\documentclass[12pt]{article}
\usepackage{amsmath,amssymb}
\newcommand{\be}{\begin{equation}}
\newcommand{\ee}{\end{equation}}
\newcommand{\bea}{\begin{eqnarray}}
\newcommand{\eea}{\end{eqnarray}}
\newcommand{\nn}{\nonumber \\}
\newcommand{\p}[1]{(\ref{#1})}
\newcommand{\lb}{\label}

\begin{document}

\begin{center}
{\Large\bf Five Lectures on Supersymmetry: \\
\vspace{0.2cm}
Elementary Introduction} \vspace{0.6cm}

{\large\bf
Evgeny Ivanov}
\vspace{0.4cm}

{\it Bogoliubov Laboratory of Theoretical Physics,
JINR, \\
141980, Dubna, Moscow Region, Russia}\\
{\tt eivanov@theor.jinr.ru}\\[8pt]
\end{center}
\vspace{0.6cm}

\begin{center}
{\it ``Symmetry in Integrable Systems and Nuclear Physics'', Tsakhkadzor, July 03 - 13, 2013}
\end{center}
\vspace{0.5cm}

\begin{abstract}
\noindent These five lectures collect elementary facts about $4D$ supersymmetric theories with emphasis on ${\cal N}=1$ supersymmetry,
as well as the basic notions of supersymmetric quantum mechanics. Contents: {\bf I.} From symmetries to supersymmetry; {\bf II.} Basic features of supersymmetry;
{\bf III.} Representations of supersymmetry; {\bf IV.} Superspace and superfields; {\bf V.} Supersymmetric quantum mechanics.
\end{abstract}

\section{Lecture I: From symmetries to supersymmetry}
\subsection{Groups and symmetries}
\noindent{\it Symmetries} play the central role in physics: They underlie all the theories of interest known to date.
Their basis is the {\it Group Theory}.
\vspace{0.3cm}

\begin{itemize}
\item \underline{Gravity}: Based on the local diffeomorphism group of the space-time, \\$Diff\, R^4$,
$x^m \Rightarrow {x^m}'(x)$.\\
\vspace{0.2cm}

\item \underline{Maxwell theory} and its non-abelian generalization, \underline{Yang-Mills theory}:
Based on the gauge groups $U(1)$ and $SU(n)$, with group parameters being arbitrary functions of the space-time point.\\
\vspace{0.2cm}

\item \underline{Standard model}, the unification of the electro-weak theory and quantum chromodynamics:
[Gauge $U(2)_{e.w.} \otimes SU(3)_c$] $\otimes$
[Global Flavor $SU(N)_f$ (broken)].\\
\vspace{0.2cm}

\item \underline{String theory}: Diffeomorphisms of the worldsheet $(z\,, \bar z)\,$.\\
\vspace{0.3cm}

\item \underline{Supergravity, Superstrings, Superbranes}: Supersymmetry (local, global, \\conformal\,, ....\,).

\end{itemize}

\noindent{\it Group}: Some manifold $G = \{g_n\}\,, \quad n = 1, 2, \ldots \,,$ such that the following axioms  are valid:
\vspace{0.2cm}

\begin{enumerate}
\item Closedness under the appropriate product:
$$
g_1 \cdot g_2 = g_3 \in G\,;
$$
\item The existence of the unit element $I \in G $:
$$
g\cdot I = I \cdot g = g\,;
$$
\item The existence of the inverse element for any $g_n \in G$:
$$
g\cdot g^{-1} = g^{-1} \cdot g = I\,;
$$
\item Associativity of the product:
$$
(g_1\cdot g_2)\cdot g_3  = g_1\cdot (g_2\cdot g_3)\,.
$$
\end{enumerate}

\noindent\underline{Simplest examples}: 1) $(1,-1)$  with respect to the standard multiplication; 2)integer numbers, with respect to
the summation, with $0$ as the unit element, etc.\\

\noindent\underline{Types of groups}: 1) finite groups; 2) infinite countable groups; 3) continuous or topological groups (Lie groups).
We will be interested in the third type.
\vspace{0.2cm}

\begin{itemize}
\item
{\bf Lie groups}:
\bea
&& G= \{g(x)\}\, \quad x :=(x^1, x^2, \ldots , x^r), \quad  r (\mbox{rank}) =  \mbox{Dim}\,G, \nn
&& g(x)\cdot g(y) = g(z(x,y)) \in G\,, \quad g(0) = I\,, \quad z(0,y) = y\,, \;z(x,0) = x\,.\nonumber
\eea
\end{itemize}
\vspace{0.2cm}

\noindent For Lie groups, one can always parametrize their elements, in a vicinity of the unit element,  as
\bea
g(x) = \exp \{x^iT_i\}\,, \quad [T_i, T_k] = c^l_{ik}T_l, \quad c^l_{ik} = -c^l_{ki}\,,\nonumber
\eea
where $T_i$ are {\it generators} and $c^l_{ik}$ are {\it structure constants}.
\vspace{0.2cm}

\noindent The generators $T_i$ span the algebra called {\it Lie algebra}. The Lie algebra is specified by its structure constants which,
in virtue of the {\it Jacobi} identity
$$
[T_l, [T_k, T_i]] + [T_i, [T_l, T_k]] + [T_k, [T_i, T_l]] = 0\,,
$$
satisfy the fundamental relation
$$
c^m_{k i} c^p_{l m} + c^m_{l k} c^p_{i m} + c^m_{i l} c^p_{k m} = 0\,.
$$
\vspace{0.2cm}

\noindent{\bf Example}: The group $SU(2)$:
\bea
&&g = \exp \{i\lambda_aT_a\}\,, \; (T_a)^\dagger = T_a\,, \quad [T_a, T_b] = i\varepsilon_{abc}T_c\,, \quad a,b,c = 1,2,3\,,\nn
&& \varepsilon_{abc}\varepsilon_{dce} +\varepsilon_{eac}\varepsilon_{bcd} + \varepsilon_{dec}\varepsilon_{acb} =0\,.\nonumber
\eea
\vspace{0.2cm}

\noindent There are two vast classes of symmetries in the Nature:

\begin{itemize}
  \item {\bf I. Internal symmetries}: Isotopic $SU(2)\,$, flavor $SU(n)\,$, etc. Their main feature:
  They are realized as transformations of
  fields without affecting the space-time coordinates. The generators are matrices acting on some external indices of fields,
  no any $x$-derivatives are present. \\{\bf Example}: Realization of $SU(2)$ on the doublet of fields
  $\psi_i(x)$ (``neutron - proton'')
\bea
&&\delta \psi_i(x) = i\lambda_a \frac{1}{2}(\sigma_a)_i^k \psi_k(x)\,, \quad [\frac{1}{2}\sigma_a,\frac{1}{2}\sigma_b] = i\varepsilon_{abc}\frac{1}{2}\sigma_c\,,\nn
&& \sigma_a \sigma_b = \delta_{ab}{\bf I} + i\varepsilon_{abc}\sigma_c\,,  \nonumber
\eea
$\sigma_a$ are Pauli matrices:
\bea
\sigma_1 = \left(\begin{array}{cc}
                  0 & 1\\
                  1 & 0
                \end{array}\right), \quad
\sigma_2 = \left(\begin{array}{cc}
                  0 & -i\\
                  i & 0
                \end{array}\right), \quad
\sigma_3 = \left(\begin{array}{cc}
                  1 & 0\\
                  0 & -1
                \end{array}\right).\nonumber
\eea

\item {\bf II. Space-time symmetries}: Lorentz, Poincar\'e and conformal groups. Generators in the realization on fields involve
$x$-derivatives.\\ {\bf Example}: Transformation of the scalar field $\varphi(x)$ in the Poincar\'e group:
\bea
&&\delta \varphi(x) :=  -i(c^mP_m  +\omega^{[mn]}L_{mn})\varphi(x) =
- c^m\partial_m  \varphi(x) - \omega^{[mn]}\frac12(x_m \partial_n - x_n\partial_m)\varphi(x) \,, \nn
&& P_m = \frac{1}{i}\partial_m\,, \quad  L_{mn} = \frac{1}{2i}(x_m \partial_n - x_n\partial_m)\,,\quad m,n = 0,1,2,3\,. \nonumber
\eea

\end{itemize}

\subsection{Invariant Lagrangians}
The primary fundamental symmetry principle is the invariance of the action:
\bea
S = \int d^4x {\cal L}(\phi_A, \partial \phi_A, \psi_\alpha, ...)\,, \quad \delta S = \frac{\delta S}{\delta \phi_A}\delta\phi_A = 0
\leftrightarrow \delta {\cal L} = \partial_m{\cal A}^m\,.\nonumber
\eea

\noindent{\bf Example}: The free Lagrangian of the scalar field
\bea
{\cal L}^{(1)}_{free} = \frac{1}{2}\partial^m \phi(x)\partial_m \phi(x) \nonumber
\eea
transforms under the Poincar\'e group as
\bea
\delta_\omega{\cal L}^{(1)}_{free} = -\frac{1}{2}\partial_n(\omega^{mn}x_m\partial^s\phi\partial_s\phi)\,, \quad
\delta_c{\cal L}^{(1)}_{free} = - \frac{1}{2} c^m\partial_m (\partial^n \phi \partial_n \phi), \nonumber
\eea
whence the invariance of the relevant action follows.\\

\noindent In the systems with few scalar fields one can realize internal symmetries. The free Lagrangian of one complex field
\bea
{\cal L}^{(2)}_{free} = \partial^m \phi(x)\partial_m \bar\phi(x) \nonumber
\eea
is invariant under $U(1)$ symmetry
\bea
\delta \phi = i \lambda \phi\,, \quad \delta \bar\phi = -i \lambda \bar\phi,\nonumber
\eea
three real scalar fields can be joined into a triplet of the group $SU(2)$:
\bea
{\cal L}^{(3)}_{free} = \frac{1}{2}\partial^m \phi_a(x)\partial_m \phi_a(x)\,, \quad
\delta \phi_a = \varepsilon_{abc}\lambda_b \phi_c \;\Rightarrow \; \delta {\cal L}^{(3)}_{free} = 0\,.\nonumber
\eea

\noindent One more possibility to construct $SU(2)$ invariant Lagrangian is to join two complex scalar fields into
$SU(2)$ doublet
\bea
&& {\cal L}^{(4)}_{free} = \partial^m \phi_\alpha (x)\partial_m \bar\phi^\alpha(x)\,, \nn
&&\delta \phi_\alpha = \frac{i}{2} \lambda_a (\sigma_a)_\alpha^\beta \phi_\beta\,, \quad
\delta \bar\phi^\alpha = -\frac{i}{2} \lambda_a (\sigma_a)^\alpha_\beta \bar\phi^\beta\,, \;\Rightarrow \;
\delta {\cal L}^{(4)}_{free} = 0\,.\nonumber
\eea
\vspace{0.2cm}

\noindent Extending the sets of fields (and adding interaction terms), we can further enlarge internal symmetries.\\
\vspace{0.4cm}

\noindent The characteristic feature of all these symmetries is that the group parameters are ordinary commuting numbers, and so the group transformations
do not mix
{\it bosonic fields} ({\it Bose-Einstein} statistics, integer spins $0, 1, \ldots$) with
{\it fermionic fields} ({\it Fermi-Dirac} statistics, half-integer spins $1/2, 3/2, \ldots$).
The bosonic and fermionic parts of the Lagrangian are invariant {\it separately}.

\subsection{Supersymmetry as symmetry between bosons and fermions}

Let us now consider a sum of the free Lagrangians of the massless complex scalar field $\varphi(x)$ and
the Weyl fermionic field $\psi^\alpha(x)$
\bea
{\cal L}_{\phi + \psi} = \partial^m \varphi \partial_m \bar\varphi -
\frac{i}{4} \Big[\psi^{\alpha}(\sigma^m)_{\alpha\dot\alpha}\partial_m\bar\psi^{\dot\alpha}
- \partial_m\psi^{\alpha}(\sigma^m)_{\alpha\dot\alpha}\bar\psi^{\dot\alpha}\Big]\,, \nonumber
\eea
where $(\sigma^m)_{\alpha\dot\alpha} = (\delta_{\alpha\dot\alpha}, (\sigma^a)_{\alpha\dot\alpha})$ are the so called sigma matrices,
the basic object of the spinor two-component formalism of the Lorentz group (they are invariant under simultaneous Lorentz transformation of
the vector $m =0, 1, 2, 3\,$, and spinor $\alpha, \dot\alpha =1, 2$ indices).\\

\noindent The evident symmetries of this Lagrangian are Poincar\'e and phase $U(1)$ symmetries which separately act on $\varphi(x)$
and $\psi^\alpha(x)$. \\

\noindent However, there is a new much less obvious symmetry. Namely, this Lagrangian transforms by a total derivative under the following transformations
mixing bosonic and fermionic fields
\bea
&& \delta \varphi = -\epsilon^\alpha\psi_\alpha \,, \quad
\delta \bar\varphi = -\bar\psi_{\dot\alpha}\bar\epsilon^{\dot\alpha} \,, \quad
\delta\psi_\alpha = 2i (\sigma^m)_{\alpha\dot\alpha}\bar\epsilon^{\dot\alpha}\partial_m\varphi\,. \nonumber
\eea
One sees that the transformation parameters $\epsilon^\alpha, \bar\epsilon^{\dot\alpha}$ have the dimension $cm^{1/2}$,
so these transformations {\it do not} define an internal symmetry (the relevant group parameters would be  {\it dimensionless}). Moreover,
for the action to be invariant, these parameters should {\it anticommute} among themselves and with the fermionic fields,
$\{\epsilon, \epsilon\} = \{\epsilon, \bar\epsilon\} = \{\epsilon(\bar\epsilon), \psi\} = 0\,$, and {\it commute} with the scalar field,
$ [\epsilon(\bar\epsilon), \varphi] = 0\,$, and with the parameters of the ordinary symmetries, e.g., $ [\epsilon(\bar\epsilon), c^m] = 0\,$. \\

\noindent To see which kind of algebraic structure is behind this invariance one needs to consider the {\it Lie bracket} of two successive
transformations on the scalar $\varphi(x)$:
\bea
(\delta_1\delta_2 - \delta_2\delta_1)\varphi = -(\epsilon_2^\alpha \delta_1\psi_\alpha) - (\epsilon_1^\alpha \delta_2\psi_\alpha)
 = 2\left(\epsilon_1 \sigma^m \bar\epsilon_2 - \epsilon_2 \sigma^m \bar\epsilon_1 \right)(\frac{1}{i}\partial_m\varphi). \nonumber
\eea
Thus the result is an infinitesimal  4-translation with the parameter
$i\left(\epsilon_1 \sigma^m \bar\epsilon_2 - \epsilon_2 \sigma^m \bar\epsilon_1 \right)$.\\

\noindent Rewriting the $\epsilon$ variation in the form
\bea
\delta \varphi = i \left(\epsilon^\alpha Q_\alpha + \bar\epsilon_{\dot\alpha}\bar Q^{\dot\alpha}\right) \varphi\,, \nonumber
\eea
and taking into account that the spinor parameters  {\it anticommute} with  $Q_\alpha, \bar Q^{\dot\alpha}$,
we find that the above {\it Lie bracket} structure is equivalent to the following
{\it anticommutation} relations for the supergenerators
\bea
&& \{Q_\alpha, \bar Q_{\dot\beta}\} = 2\,(\sigma^m)_{\alpha\dot\beta}P_m\,, \quad P_m = \frac{1}{i} \frac{\partial}{\partial x^m}\,,  \nn
&& \{Q_\alpha, Q_{\beta}\} =  \{\bar Q_{\dot\alpha}, \bar Q_{\dot\beta}\} = 0\,, \nn
&&[P_m, Q_\alpha] = [P_m, \bar Q_{\dot\alpha}] = 0\,. \nonumber
\eea
This is what is called ${\cal N}=1$ {\it Poincar\'e superalgebra}.

\section{Lecture II: Basic features of supersymmetry}

The full set of the (anti)commutation relations of the ${\cal N}=1$ Poincar\'e superalgebra reads
\bea
&& \{Q_\alpha, \bar Q_{\dot\beta}\} = 2\,(\sigma^m)_{\alpha\dot\beta}P_m\,, \nn
&& \{Q_\alpha, Q_{\beta}\} =  \{\bar Q_{\dot\alpha}, \bar Q_{\dot\beta}\} = 0\,, \nn
&&[ P_m, Q_\alpha] = [P_m, \bar Q_{\dot\alpha}] = 0\,,\lb{N1c} \\
&& [J_{mn}, Q_\alpha] = -\frac{1}{2}\,(\sigma_{mn})_\alpha^{\;\beta} Q_\beta\,, \quad
[J_{mn}, \bar Q_{\dot\alpha}] =  \frac{1}{2}\,(\tilde{\sigma}_{mn})_{\dot\alpha}^{\;\dot\beta}
\bar Q_{\dot\beta}\,, \nn
&& [J_{mn}, P_s] = i\left(\eta_{ns}P_m - \eta_{ms} P_n\right), \nn
&& [J_{mn}, J_{sq}] = i\left(\eta_{ns}J_{mq} - \eta_{ms} J_{nq} +\eta_{nq}J_{sm} - \eta_{mq} J_{sn}\right), \nn
&& [R, Q_\alpha] = Q_\alpha\,, \quad [R, \bar Q_{\dot\alpha}] = -\bar Q_{\dot\alpha}\,
\quad [R, P_m] = [R, J_{mn}] = 0\,. \nonumber
\eea
Here, $J_{mn} = L_{mn} + S_{mn}$ are the full Lorentz group generators ($S_{mn}$ is the {\it spin} part acting on the external vector and spinor indices)
and $R$ is the generator of an extra internal $U(1)$
symmetry (the so-called $R$ symmetry). Also,
\bea
&& \eta_{mn} = \mbox{diag}(1,-1,-1,-1)\,, \quad (\sigma_{mn})_\alpha^\beta
= \frac{i}{2}\left(\sigma^m \tilde{\sigma}^n - \sigma^n\tilde{\sigma}^m\right)_\alpha^\beta\,, \nn
&& (\tilde{\sigma}^{mn})_{\dot\alpha}^{\dot\beta} =
\frac{i}{2}\left(\tilde{\sigma}^m \sigma^n - \tilde{\sigma}^n\sigma^m\right)_{\dot\alpha}^{\dot\beta}\,, \quad
\tilde{\sigma}^m{}^{\dot\alpha\alpha} = (\delta^{\dot\alpha\alpha}, -\sigma^a{}^{\dot\alpha\alpha})\,. \nonumber
\eea

\noindent Some important common features and consequences of supersymmetry  can be figured out just from these (anti)commutation relations.

\begin{itemize}
\item The Poincar\'e superalgebra is an example of $Z_2$-graded algebra. The latter is defined in the following way:
one ascribes parities $\pm 1$ to all its elements, calling them, respectively, {\it even} (parity $+1$) and {\it odd} (parity $-1$) elements,
and requires the structure relations to respect these parities:
\bea
&& [\mbox{odd}, \mbox{odd}] \sim \mbox{even}\,, \; [\mbox{even}, \mbox{odd}] \sim \mbox{odd}\,, \;
[\mbox{even}, \mbox{even}] \sim \mbox{even}\,.\nonumber
\eea
{}From the above (anti)commutation relations we observe that the spinor generators $Q_\alpha, \bar Q_{\dot\alpha}$ can be assigned the parity -1
and so they are {\it odd}; all bosonic generators can be assigned the parity +1 and so they are {\it even}.
\item
Lie superalgebras satisfy the same axioms as the Lie algebras, the difference is that the relevant generators satisfy the
{\it graded} Jacobi identities, because the fermionic generators are subject to the {\it anticommutation} relations. E.g.,
\bea
&& \{[B_1, F_2], F_3\} - \{[F_3, B_1], F_2\} + [\{F_2, F_3\}, B_1] = 0\,, \nn
&& [\{F_1, F_2\}, F_3]  + [\{F_3, F_1\}, F_2] + [\{F_2, F_3\}, F_1] = 0\,,  \nonumber
\eea
where $B_1$ is a bosonic generator and $F_1, F_2, F_3$ are fermionic ones.

\item Since the generators $Q_\alpha, \bar Q_{\dot\alpha}$ are fermionic, irreducible  multiplets of supersymmetry ({\it supermultiplets})
should unify bosons with fermions. Action of the spinor generators on the bosonic state yields a fermionic state and vice versa.

\item
Since the translation operator $P_m$ is non-vanishing on {\it any} field given on the Minkowski space,
the same should be true for the spinor generators as well. So any field should belong to
a non-trivial supermultiplet.

\item
It follows from the relations $[ P_m, Q_\alpha] = [P_m, \bar Q_{\dot\alpha}] = 0$ that
$[ P^2, Q_\alpha] = [P^2, \bar Q_{\dot\alpha}] = 0$. The operator $P^2 = P^mP_m$ is a Casimir of the Poincar\'e group,
$P^2=m^2$. So it is also a Casimir of the Poincar\'e supergroup. Hence all components of the
irreducible supermultiplet should have the same mass. No mass degeneracy between bosons and fermions is observed in Nature,
so supersymmetry should be {\it broken} in one or another way.

\item
In any representation of supersymmetry, such that the operator $P_m$ is invertible, there should be {\it equal numbers}
of bosons and fermions.

\item
In any supersymmetric theory the energy $P_0$ should be non-negative. Indeed, from the basic anticommutator it follows
\bea
\sum_{\alpha =1, 2} \left(|Q_\alpha|^2 + |\bar Q_{\dot\alpha}|^2\right) = 4P_0 \geq 0\,.\nonumber
\eea
\item
Rigid supersymmetry, with constant parameters,  implies the translation invariance. Gauge supersymmetry, with the parameters being
arbitrary functions of the space-time point, implies the invariance under arbitrary {\it diffeomorphisms} of the Minkowski space.
Hence the theory of gauged supersymmetry necessarily contains {\it gravity}. The theory of gauged supersymmetry  is {\it supergravity}.
Its basic gauge fields are {\it graviton} (spin 2) and {\it gravitino} (spin 3/2).

\end{itemize}

\subsection{Extended supersymmetry}

\noindent Supersymmetry allows one to evade the famous {\it Coleman-Mandula} theorem about impossibility of non-trivial unification
of the space-time symmetries with
the internal ones. It states that any symmetry of such type (in dimensions $\geq 3$), under the standard assumptions
about the spin-statistics relation,
is inevitably reduced to the direct product of the Poincar\'e group and the internal symmetry group.
\vspace{0.1cm}

\noindent The arguments of this theorem do not apply to superalgebras, when one deals with both commutation and anticommutation relations.
{\bf Haag, Lopushanski}, and {\bf Sohnius} showed that the most general superextension of the Poincar\'e group algebra is given
by the following relations
\bea
&& \{Q_\alpha^i, \bar Q_{\dot\beta k}\} = 2\delta^i_k\,(\sigma^m)_{\alpha\dot\beta}P_m\,, \nn
&& \{Q_\alpha^i, Q_{\beta}^j\} = \epsilon_{\alpha\beta} Z^{ij}\,, \quad  \{\bar Q_{\dot\alpha i}, \bar Q_{\dot\beta j}\}
=  \epsilon_{\dot\alpha\dot\beta} \bar{Z}_{ij}\,, \nn
&& [T^i_j, Q^k_\alpha] = -i\left(\delta^k_j Q^i_\alpha - \frac{1}{{\cal N}} \delta^i_j Q^k_\alpha\right), \quad
[T^i_j, \bar Q_{\dot\alpha k}] = i\left(\delta^i_k \bar Q_{\dot\alpha j}
- \frac{1}{{\cal N}} \delta^i_j \bar Q_{\dot\alpha k}\right),\nn
&& [T^i_j, T_l^k] = i\left(\delta^i_l T^k_j - \delta^k_j T^i_l\right), \nonumber
\eea
where  $T^i_j$ ($(T^i_j)^\dagger = - T_i^j\,, \;T^i_i = 0$) are generators of the group $SU({\cal N})\,$.
The generators $Z^{ij} = -Z^{ji}, \bar Z_{ij} = -\bar Z_{ji}$ are central charges, they commute with all generators except the
$SU({\cal N})$ ones
\bea
[Z, Z] = [Z, \bar Z] = [Z, P] = [Z, J] = [Z, Q] = [Z, \bar Q] = 0\,.\nonumber
\eea
The relevant supergroup is called ${\cal N}${\it -extended Poincar\'e supergroup}.\\

\noindent Due to the property that the spinor generators $Q_\alpha^i, \bar Q_{\dot\beta k}$ carry the internal symmetry indices, the
supermultiplets of extended supersymmetries join fields having not only different statistics and spins, but also belonging to different
representations of the internal symmetry group $U({\cal N})$. In other words, in the framework of extended supersymmetry
the {\it actual unification} of the space-time and internal symmetries comes about. The relevant supergravities can involve, as a subsector, gauge
theories of internal symmetries, i.e. yield non-trivial unifications of {\it Einstein} gravity with {\it Yang-Mills} theories.

\subsection{Auxiliary fields}
An important ingredient of supersymmetric theories is the {\it auxiliary fields}. They ensure the closedness
of the supersymmetry transformations {\it off mass shell}.
\vspace{0.2cm}

\noindent Let us come back to the realization of ${\cal N}=1$ supersymmetry on the fields $\varphi(x), \psi_\alpha(x)$ and
calculate Lie bracket of the odd transformations on $\psi_\alpha(x)$:
\bea
(\delta_1\delta_2 - \delta_2\delta_1)\psi_\alpha =  -2i
\left(\epsilon_1 \sigma^m \bar\epsilon_2 - \epsilon_2 \sigma^m \bar\epsilon_1 \right) \partial_m\psi_\alpha
+2i \left[ \epsilon_{1\alpha}\bar\epsilon_{2 \dot\alpha}(\tilde{\sigma}^m)^{\dot\alpha\beta}\partial_m\psi_\beta
- (1 \leftrightarrow 2)\right].\nonumber
\eea
The first term in the r.h.s. is the translation one, as for $\varphi(x)$. However, there is one extra term. It is clear that
the Lie bracket should have the same form on all members of the supermultiplet, i.e. reduce to translations. The condition of vanishing
of the second term is
\bea
\tilde{\sigma}^m\partial_m\psi = \sigma^m\partial_m\bar\psi = 0\,.\nonumber
\eea
But this is just the free equation of motion for $\psi_\alpha(x)$. Thus ${\cal N}=1$ supersymmetry is closed only on-shell, i.e.
modulo equations of motion.
\vspace{0.2cm}

\noindent How to secure the off-shell closure? The way out is to introduce a new field $F(x)$ of
non-canonical dimension $cm^{-2}$ and to extend the free action of $\varphi, \psi_\alpha$ as
\bea
{\cal L}_{\phi + \psi +F} = \partial^m \varphi \partial_m \bar\varphi -
\frac{i}{4} \Big[\psi^{\alpha}(\sigma^m)_{\alpha\dot\alpha}\partial_m\bar\psi^{\dot\alpha}
- \partial_m\psi^{\alpha}(\sigma^m)_{\alpha\dot\alpha}\bar\psi^{\dot\alpha}\Big] + F\bar F\,. \nonumber
\eea
It is invariant, up to a total derivative, under the modified transformations having the correct closure for all fields:
\bea
\delta \phi = -\epsilon^\alpha\psi_\alpha \,, \;
\delta\psi_\alpha = -2i (\sigma^m)_{\alpha\dot\alpha}\bar\epsilon^{\dot\alpha}\partial_m\phi - 2\epsilon_\alpha F\,, \;
\delta F = -i\bar\epsilon^{\dot\alpha}(\sigma^m)_{\alpha\dot\alpha}\partial_m \psi^\alpha\,. \label{Off}
\eea

\noindent The auxiliary fields satisfy the {\it algebraic} equations of motion
\bea
F = \bar F = 0\,.\nonumber
\eea
After substitution of this solution back in the Lagrangian and supersymmetry transformations, we reproduce the previous on-shell realization.
The auxiliary fields do not propagate also in the quantum case, possessing delta-function propagators.
\vspace{0.2cm}

\noindent The only (but very important!) role of the auxiliary fields is just to ensure the correct off-shell realization of supersymmetry,
such that it does not depend on the precise choice of the invariant Lagrangian, like in the cases of ordinary symmetries.
\vspace{0.2cm}

\noindent The simplest non-trivial choice of the Lagrangian is as follows
\bea
&& {\cal L}_{\mbox{wz}} = \partial^m \phi \partial_m \bar\phi -
\frac{i}{4} \left[\psi^{\alpha}(\sigma^m)_{\alpha\dot\alpha}\partial_m\bar\psi^{\dot\alpha}
- \partial_m\psi^{\alpha}(\sigma^m)_{\alpha\dot\alpha}\bar\psi^{\dot\alpha}\right] + F\bar F \nn
&& + \,\left[ m\left(\phi F - \frac{1}{4}\psi\psi\right) + g\left(\phi^2 F - \frac{1}{2}\phi \psi\psi\right) + c.c. \right]. \nonumber
\eea
This model was the first example of {\it renormalizable} supersymmetric quantum field theory and it is called the
{\bf Wess-Zumino} model, after names of its discoverers. The Lagrangian ${\cal L}_{\mbox{wz}}$ is invariant
under the same transformations as the free Lagrangian we have considered before.\\

\noindent The Wess-Zumino model  Lagrangian was originally found by the ``trying and error''
method. The systematic way of constructing invariant off-shell Lagrangians is the {\it superfield} method which we will discuss in the Lectures IV and V.
\vspace{0.2cm}

\noindent Using this systematic method, one can equally construct more general Lagrangians of the fields $(\phi, \psi_\alpha, F)$,
invariant under the same linear {\it off-shell} ${\cal N}=1$ supersymmetry transformations \p{Off}. After eliminating the auxiliary fields from these
Lagrangians by their equations of motion, we will obtain the Lagrangians in terms of the physical fields $(\phi, \psi_\alpha)$ only.
These physical Lagrangians are invariant under the nonlinear {\it on-shell} ${\cal N}=1$ supersymmetry transformations
the precise form of which depends on the form of the on-shell Lagrangian, though it is uniquely specified by the off-shell Lagrangian.
\vspace{0.4cm}

\noindent To summarize, the fields $(\phi, \psi_\alpha, F)$ form the set {\it closed} under the off-shell ${\cal N}=1$ supersymmetry
transformations, and it is impossible to select any lesser closed set of fields in it. Thus these fields constitute the simplest irreducible
multiplet of ${\cal N}=1$ supersymmetry. It is called {\it scalar} ${\cal N}=1$ {\it supermultiplet}.

\section{Lecture III: Representations of supersymmetry}

The fields on Minkowski space are distributed over the irreducible multiplets of the Poincar\'e group  according
to the eigenvalues of two Casimirs of this group: the square of $P_m$ (which is $m^2$) and the square of the Pauli-Lubanski vector (which $\propto s(s+1)$, where $s$
is the spin of the field).
For the case of zero mass the diverse Poincar\'e group multiplets  are characterized by the {\it helicity}, the projection of spin on the
direction of motion. What about irreps of supersymmetry? Once again, the contents of the supermultiplets are different
for massive and massless cases.

\subsection{Massive case}
Choose the rest frame
\bea
P_m = (m, 0, 0, 0)\,. \nonumber
\eea
In this frame
\bea
\mbox{(a)} \; \{Q_\alpha, Q_\beta \} = \{\bar Q_{\dot\alpha}, \bar Q_{\dot\beta} \} = 0\,; \quad
\mbox{(b)} \;\{Q_\alpha, \bar Q_{\dot\beta} \} = 2m \delta_{\alpha\dot\beta}\,, \nonumber
\eea
i.e. ${\cal N}=1$ superalgebra becomes the Clifford algebra of two mutually conjugated fermionic creation and destruction operators.
$\bar Q_{\dot\alpha} $ and  $Q_\alpha $. Define the ``Clifford vacuum'' $|s>$ as the irrep of the
Poincar\'e group with mass $m$ and spin $s$:
\bea
Q_\alpha |s> = 0\,. \nonumber
\eea

\noindent An irrep of the full supersymmetry can be then produced by the successive action of $\bar Q_{\dot\alpha}$ on the vacuum $|s>$:
\bea
\begin{array}{ccc} {\phantom{mm}\rm State\phantom{mm}}&{\phantom{mm}
\rm Spin \phantom{mm}}&\phantom{mm}\mbox{\# of  components}\phantom{mm}\\
\begin{array}{r}\vert s\rangle \\ \bar Q_{\dot\alpha}\vert s\rangle \\ (\bar Q)^2\vert s\rangle \end{array}
&\begin{array}{c}s\\ s\pm 1/2\\s \end{array} &
\begin{array}{c}2s+1\\ 4s+2\\ 2s+1\end{array}
\end{array}
\nonumber
\eea
Here $(\bar Q)^2 \equiv \bar Q_{\dot\alpha}\bar Q^{\dot\alpha}$. Further acting by $\bar Q_{\dot\alpha}$ yields zero.
Thus the full number of states is $2^2(2s+1)$, one half being fermions and the second one bosons.
The dimensionality of the Clifford vacuum (the number of independent states in it) is  just $d_{|s>}=2s+1$.
\vspace{0.1cm}

\noindent Since off shell $P^2 \neq 0$, this spin contents characterizes any off-shell supermultiplet.
E.g., the scalar multiplet corresponds to $s=0$: In this case $s+1/2 = 1/2$ and we are left just
with two complex scalars and one Weyl fermion.
\vspace{0.1cm}

\noindent Thus massive ${\cal N}=1$ supermultiplets are entirely specified by the spin $s$ of their Clifford vacua.
This spin is called {\it superspin} $Y$ of the given ${\cal N}=1$ supermultiplet.
Each multiplet with $P^2 \neq 0$ and superspin $Y$ involves the following set of spins
\bea
Y\,, \; Y + \frac{1}{2}, \;  Y - \frac{1}{2}\,, \; Y\,. \nonumber
\eea

\noindent The {\it scalar} supermultiplet ($Y=0$) contains spins $1/2, (0)^2$ and describes ${\cal N}=1$ matter.
The supermultiplet with  $Y=1/2$ involves states with spins $1, (1/2)^2, 0$ and stands for the {\it gauge}
supermultiplet. The supermultiplet with $Y = 3/2$ has the spin content  $(3/2)^2, 2, 1$. It is the so-called ${\cal N}=1$
{\it Weyl} supermultiplet. It corresponds to conformal ${\cal N}=1$ supergravity.

\subsection{Massless case}
We can choose the frame
\bea
P_m = (p, 0, 0, p)\,, \quad P^mP_m = 0\,.\nonumber
\eea
The only non-zero anticommutator in this frame is
\bea
\{Q_\alpha, \bar Q_{\dot\beta} \} = 2 \,p\,(I + \sigma^3)_{\alpha\dot\alpha}\,.\nonumber
\eea
The full set of the anticommutation relations is
\bea
&& \{Q_1, \bar Q_{\dot{1}} \} = 4 p\,, \quad  \{Q_1, \bar Q_{\dot{2}} \} = \{Q_2, \bar Q_{\dot{2}} \} =
\{Q_2, \bar Q_{\dot{1}} \} = 0\,, \nn
&& \{Q_\alpha, Q_\beta \} = \{\bar Q_{\dot\alpha}, \bar Q_{\dot\beta} \} = 0\,.\nonumber
\eea
Then one can define the Clifford vacuum $|\lambda>$ with the helicity
$\lambda $ by the conditions
\bea
Q_1|\lambda> = Q_2|\lambda> = \bar{Q}_{\dot{2}}|\lambda> = 0\,.\nonumber
\eea

\noindent The only creation operator is $\bar Q_{\dot{1}}$. Due to its nilpotency, $(\bar Q_{\dot{1}})^2 = 0$,
the procedure of constructing the irreducible set of states terminates at the 1st step:
\bea
\begin{array}{ccc} {\phantom{mm}\rm State\phantom{mm}}&{\phantom{mm}\rm Helicity \phantom{mm}}&\phantom{mm}
\mbox{\# of components}\phantom{mm}\\
\begin{array}{r}\vert \lambda\rangle \\ \bar Q_{\dot 1}\vert \lambda\rangle
\end{array}
&\begin{array}{c}\lambda\\ \lambda - 1/2 \end{array}
&\begin{array}{c}1\\ 1\end{array}
\end{array}
\nonumber
\eea

\noindent Thus in ${\cal N}=1$ supersymmetry the massless supermultiplets are formed by pairs of states with the adjacent helicities,
$\vert\lambda\rangle$, $\vert\lambda -1/2\rangle$.
In particular, massless particle with zero helicity should be accompanied by a particle with the
helicity $-1/2$, a particle with $\lambda = 1/2$ should be paired with a particle having $\lambda = 0$,
helicities $\pm 1$ can be embedded  either into the multiplets $(1, 1/2)$, $(-1/2, -1)$, or
$(-1, -3/2)$, $(3/2, 1)$, the minimal embeddings for the helicities  $\pm 2$ are into the multiplets
$(2, 3/2)$ and $(-3/2, -2)$, etc. The multiplets with the opposite helicities are related through  {\it CPT}
conjugation.

\subsection{Massless multiplets of ${\cal N}$ extended supersymmetry}
In this case (without central charges) the only non-vanishing anticommutator is
\bea
\{Q_1^i, \bar Q_{\dot{1}j} \} = 4 \delta^i_j p\,.
\eea
The Clifford vacuum $|\lambda>$ is defined by
\bea
Q^i_1\vert\lambda\rangle =Q^i_2\vert\lambda\rangle =\bar Q_{\dot{2} i}\vert\lambda\rangle =0\,,
\eea
and the irreducible tower of states is constructed by acting on the vacuum by ${\cal N}$ independent creation operators
$\bar Q_{\dot{1} i}$:
\bea
\begin{array}{ccc} {\phantom{mm}\rm State\phantom{mm}}&{\phantom{mm}\rm Helicity\phantom{mm}}
&\phantom{mm}\mbox{\# of components}\phantom{mm}\\
\begin{array}{r}\vert \lambda\rangle \\ \bar Q_{\dot 1i}\vert \lambda\rangle \\
\bar Q_{\dot 1i}\bar Q_{\dot 1j}\vert \lambda\rangle\\ \vdots\phantom{\vert
\lambda\rangle}\\ (\bar Q)^{{\cal N}}\vert\lambda\rangle \end{array}
&\begin{array}{c}\lambda\\ \lambda - 1/2\\ \lambda -1\\ \vdots\\ \lambda -{\cal N}/2 \end{array}
&\begin{array}{c}1\\ {\cal N}\\ {\cal N}({\cal N}-1)/2\\ \vdots\\ 1\end{array}
\end{array}
\nonumber
\eea
For ${\cal N}=2$ supersymmetry, irreps are formed by the states $\vert\lambda\rangle$,
$\vert\lambda -1/2\rangle^2$, $\vert\lambda -1\rangle$, etc.\\

\noindent Recall that the multiplets with opposite
helicities can be obtained via {\it CPT} conjugation. Of special interest are the so-called  {\it ``self-conjugated''} multiplets which,
from the very beginning, involve the full spectrum of helicities from $\lambda $ to $-\lambda$. Equating
\bea
\lambda -{\cal N}/2 = -\lambda \; \Rightarrow \lambda = {\cal N}/4\,,
\eea
we find that, up to ${\cal N}= 8$, there exist the following self-conjugated massless supermultiplets
\bea
   & & \mbox{${\cal N}=2$ matter multiplet:}\ \ \ \ \ \ \ \ \ \ \ \ \ \ \  1/2,(0)^2,-1/2;  \nn
   & & \mbox{${\cal N}=4$ gauge multiplet:}\ \ \ \ \ \ \ \ \ \ \ \ \ \ \ 1,(1/2)^4,(0)^6,(-1/2)^4,-1; \nn
   & & \mbox{${\cal N}=8$ supergravity multiplet:}\ \ \ \ \ \  2,(3/2)^8,(1)^{28},(1/2)^{56},(0)^{70},\nn
      & & \phantom{\mbox{${\cal N}=8$ supergravity multiplet:}}\ \ \ \ \ \ (-1/2)^{56},(-1)^{28},(-3/2)^8,-2.
\nonumber
\eea
Note that for ${\cal N}>8$ the massless supermultiplets would  include helicities $> 2$.
The relevant theories are called ``higher-spin theories'' and, for self-consistency at the full interaction level,
they should include the whole infinite set of such spins (helicities). Such complicated theories are under intensive study at present,
but their consideration is beyond the scope of my lectures.

\section{Lecture IV: Superspace and superfields}
\subsection{Superspace}
When considering one or another symmetry and constructing physical models invariant with respect to it,
it is very important to find out the proper space and/or the
fundamental multiplet on which this symmetry is realized in the most
natural and simplest way.\\

\noindent The Poincar\'e group has a natural
realization in the Minkowski space $x^m, m=0,1,2,3\,,$ as the group of
linear rotations and shifts of $x^m$ preserving the flat invariant
interval $ds^2 = \eta_{mn}dx^m dx^n$. Analogously, {\it
supersymmetry} has a natural realization in the Minkowski
{\it superspace}. \\

\noindent The translation generators $P_m$ can be realized as shifts of  $x^m, {x^m}' = x^m + c^m$. In the case of
${\cal N}=1$ supersymmetry we have additional spinor
generators $Q_\alpha, \bar Q_{\dot\alpha}$ and
anticommuting parameters $\epsilon^\alpha,
\bar\epsilon^{\dot\alpha}$. Then it is natural to introduce new
spinor coordinates $\theta^\alpha,
\bar\theta^{\dot\alpha}$ having the same dimension
$cm^{1/2}$ as the spinor parameters and to realize the
spinorial generators as shifts of these new coordinates
\bea
\theta^\alpha{}' = \theta^\alpha +\epsilon^\alpha\,,\quad
\bar\theta^{\dot\alpha}{}' = \bar\theta^{\dot\alpha}
+\bar\epsilon^{\dot\alpha}\,. \nonumber
\eea

\noindent The extended  manifold
\bea {\cal M}^{(4|4)} = \left( x^m\,,
\;\theta^\alpha\,, \;\bar\theta^{\dot\alpha}\right),\nonumber
\eea
is called ${\cal N}=1$ {\bf Minkowski superspace}.\\

\noindent Its natural generalization is
\bea
{\cal M}^{(4|4{\cal N})} = \left(
x^m\,,\;\theta^\alpha_i\,, \;\bar\theta^{\dot\alpha\,i}\right),
\nonumber
\eea
and it is called ${\cal N}$ {\bf extended Minkowski superspace}.\\

\noindent The spinor coordinates are called odd or Grassmann coordinates
 and have the Grassmann parity $-1$, while $x^m$ are even coordinates having the Grassmann parity $+1$
\bea
[\theta^\alpha_i, x^m] =
[\bar\theta^{\dot\alpha\,i}, x^m] = 0\,, \quad \{\theta^\alpha_i,
\theta^\beta_k\} = \{\theta^\alpha_i, \bar\theta^{\dot\beta\,k} \} =
0\,. \nonumber
\eea
The spinor coordinates also anticommute with
the parameters $\epsilon^\alpha,
\bar\epsilon^{\dot\alpha}$.\\

\noindent Since two supertranslations yield a shift of $x^m$, they should be non-trivially realized on $x^m$.
In the ${\cal N}=1$  case:
\bea
x^m{}'  =
x^m -i(\epsilon\sigma^m \bar\theta- \theta\sigma^m\bar\epsilon)\,,
\quad (\delta_1\delta_2 - \delta_2\delta_1)x^m =
2i(\epsilon_1\sigma^m \bar\epsilon_2 -
\epsilon_2\sigma^m\bar\epsilon_1)
\nonumber
\eea
(an analogous
transformation takes place in the general case of
${\cal N}$ extended supersymmetry).

\subsection{Superfields}
{\it Superfields} are functions on superspace, such that they have
definite transformation properties under supersymmetry. The general
scalar ${\cal N}=1$ superfield is $\Phi(x, \theta, \bar\theta)$ with the following
transformation law
\bea
\Phi'(x', \theta', \bar\theta')
= \Phi(x, \theta, \bar\theta)\,.\nonumber
\eea

\noindent The most important property of superfield is that its series expansion in Grassmann coordinates terminates at the finite step.
The reason is that these coordinates are {\it nilpotent}, because they
anticommute. E.g., $\{\theta_\alpha, \theta_\beta\} = 0
\Rightarrow \theta_1\theta_1 = \theta_2\theta_2 = 0\,$.
Then
\bea
\Phi(x, \theta, \bar\theta) &=& \phi(x) +
\theta^\alpha\,\psi_\alpha(x) +
\bar\theta_{\dot\alpha}\,\bar\chi^{\dot\alpha}(x)+ \theta^2\,M(x) +
\bar\theta^2\,N(x) \nn
&&+\, \theta \sigma^m\bar\theta\,A_m(x) +
\bar\theta^2 \,\theta^\alpha \,\rho_\alpha(x) +
\theta^2\,\bar\theta_{\dot\alpha} \,\bar\lambda^{\dot\alpha}(x) +
\theta^2\,\bar\theta^2\, D(x)\,,  \nonumber
\eea
where $\theta^2 :=\theta^\alpha\theta_\alpha=\epsilon_{\alpha\beta}\theta^\alpha\theta^\beta\,,
\;\bar\theta^2  = \bar\theta_{\dot\alpha}\bar\theta^{\dot\alpha} =
\epsilon_{\dot\alpha\dot\beta}\bar\theta^{\dot\beta}\bar\theta^{\dot\alpha}\,,
\;\epsilon_{12} = \epsilon_{\dot{1}\dot{2}} =1$.\\

\noindent Here one deals with the set of  {8} bosonic and {8} fermionic independent complex component fields.
The reality condition
\bea
\overline{(\Phi)} = \Phi \nonumber
\eea
implies
the following reality conditions for the component fields
\bea
&& \phi(x)= \overline{\phi(x)}\,, \;\;
\bar\chi_{\dot\alpha}(x) = \overline{\psi_{\alpha}(x)}\,, \;\; M(x)
= \overline{N(x)}\,, \;\; A_m(x) = \overline{A_m(x)}\,, \nn
&&
\bar\lambda^{\dot\alpha}(x) = \overline{\rho^{\alpha}(x)}\,, \;\;
D(x) = \overline{D(x)}\,.\nonumber
\eea
They leave in $\Phi$ just $(8 + 8)$ independent real
components.\\

\noindent The transformation law $\Phi'(x, \theta, \bar\theta) = \Phi(x -\delta x, \theta -\epsilon, \bar\theta - \bar\epsilon)$
implies
\bea
&& \delta \Phi = -\epsilon^\alpha
\frac{\partial \Phi}{\partial \theta^\alpha} -
\bar\epsilon_{\dot\alpha}\frac{\partial \Phi}{\partial
\bar\theta_{\dot\alpha}} - \delta x^m \frac{\partial \Phi}{\partial
x^m} \equiv i \left(\epsilon^\alpha Q_\alpha +
\bar\epsilon_{\dot\alpha}\bar Q^{\dot\alpha}\right)\Phi\,,  \nn
&&Q_\alpha = i {\partial\over
\partial \theta^\alpha } +\bar\theta^{\dot\alpha}(\sigma^m)_{\alpha\dot\alpha}
 {\partial\over \partial x^m}~, \quad
\bar Q_{\dot \alpha} = -i{\partial\over \partial
\bar\theta^{\dot\alpha}} -\theta^\alpha
(\sigma^m)_{\alpha\dot\alpha} {\partial\over \partial x^m} ~, \nn
&&
\{Q_\alpha, \bar Q_{\dot \alpha}\} = 2(\sigma^m)_{\alpha\dot\alpha}P_m\,, \; \{Q_\alpha,
Q_{\beta}\} = \{\bar Q_{\dot \alpha}, \bar Q_{\dot \beta}\} = 0\,,
\quad P_m = {1\over i}{\partial\over \partial x^m}~. \nonumber
\eea

\noindent The relevant component transformations are read off from the formula
$\delta \Phi = \delta\phi +
\theta^\alpha\delta\psi_\alpha + \ldots + \theta^2 \bar\theta^2
\delta D$.
They are
\bea
&& \delta \phi = -\epsilon
\psi -\bar\epsilon \bar\chi\,, \quad \delta \psi_\alpha = -i
(\sigma^m\bar\epsilon)_\alpha \partial_m\phi -2\epsilon_\alpha M -
(\sigma^m\bar\epsilon)_\alpha A_m\,, \ldots\,,\nn
&& \delta D =
\frac{i}{2} \partial_m \rho\sigma^m \bar\epsilon - \frac{i}{2}
\epsilon\sigma^m \partial_m\bar\lambda\,. \nonumber
\eea

\noindent These transformations uniformly close on $x^m$ translations without use of any dynamical equations. However,
the supermultiplet of fields encompassed by $\Phi(x, \theta, \bar\theta)$ is {\it reducible}: it contains in
fact both the {\it scalar} and {\it gauge}
${\cal N}=1$ supermultiplets (superspins
$Y=0$ and $Y=1/2$). How to describe
{\it irreducible} supermultiplets in the superfield
language?\\

\noindent An important element of the superspace formalism are spinor covariant derivatives
\bea
&& D_\alpha = \frac{\partial}{\partial
\theta^\alpha}+
i\bar\theta^{\dot\alpha}(\sigma^m)_{\alpha\dot\alpha}
\frac{\partial}{\partial x^m}, \quad \bar D_{\dot\alpha} =
-\frac{\partial}{\partial \bar\theta^{\dot\alpha}}- i\theta^\alpha
(\sigma^m)_{\alpha\dot\alpha} \frac{\partial}{\partial x^m}\,,\nn
&&\{D_\alpha, \bar D_{\dot\alpha} \} = -2i
(\sigma^m)_{\alpha\dot\alpha}\partial_m\,, \quad \{D_\alpha,
D_{\beta} \} =\{\bar D_{\dot\alpha}, \bar D_{\dot\beta} \} = 0\,.
\nonumber
\eea

\noindent The covariant spinor derivatives {\it anticommute} with supercharges, $\{D, Q\}  = \{D, \bar Q\} = 0$,
so $D_\alpha\Phi $  and $\bar D_{\dot\alpha}\Phi$ are again superfields, e.g.,
$$
\delta
D_\alpha\Phi = D_\alpha \delta\Phi = D_{\alpha}i
\left(\epsilon^\alpha Q_\alpha + \bar\epsilon_{\dot\alpha}\bar
Q^{\dot\alpha}\right)\Phi = i \left(\epsilon^\alpha Q_\alpha +
\bar\epsilon_{\dot\alpha}\bar Q^{\dot\alpha}\right)D_\alpha\Phi\,.
$$

\noindent Now, it becomes possible to define the ``irreducible'' superfields. ({\bf Analogy}: In Minkowski
space the vector field $A_m$
is known to carry two Poincar\'e spins $1$ and
$0$. The irreducible components are distinguished by
imposing on $A_m$ the supplementary differential
conditions
$$
\partial^mA_m = 0 \leftrightarrow \mbox{spin\; 1}\,, \quad \partial_mA_n - \partial_nA_m = 0 \leftrightarrow \mbox{spin \;0}\,.)
$$

\noindent Analogous conditions can be imposed on the superfield $\Phi$ in order to single out the irreducible multiplets
with the superspins $0$ and $1/2$. These
conditions are defined with the help of the covariant spinor
derivatives.\\

\noindent The simplest condition of this type is the {\it chirality} or {\it anti-chirality} conditions
\bea
(a)\;\bar D_{\dot\alpha}\Phi_L
(x,\theta,\bar\theta)=0\,, \quad \mbox{or} \quad
(b)\;D_{\alpha}\Phi_R (x,\theta,\bar\theta)=0\,.
\nonumber
\eea

\noindent Eq. $(a)$, e.g.,  implies
\bea
&&\Phi_L(x, \theta, \bar\theta) = \varphi_L
(x_L,\theta) = \phi(x_L)+ \theta^\alpha\psi_\alpha(x_L)+
\theta\theta F(x_L)\,, \nn
&& x_L^m=x^m+i\theta
\sigma^m\bar\theta\,,\nonumber
\eea
{\it i.e.} we are left with the
independent fields $\phi, \psi_\alpha, F$.\\

\noindent {}From the general transformation laws
of the component fields it follows that this set is closed under
${\cal N}=1$ supersymmetry:
\bea
\delta
\phi = -\epsilon \psi\,, \quad \delta \psi_\alpha = -2i
(\sigma^m\bar\epsilon)_\alpha \partial_m\phi -2\epsilon_\alpha F
\,,\quad \delta F = -i \bar\epsilon\tilde{\sigma}^m
\partial_m\psi\,. \nonumber
\eea
These are just the transformation laws of the scalar ${\cal N}=1$ supermultiplet.\\

\noindent {\it The geometric interpretation}: The coordinate set $(x_L^m, \theta^\alpha)$ is closed under ${\cal N}=1$ supersymmetry:
\bea
\delta x_L^m = 2i
\theta\sigma^m\bar\epsilon\,,\quad \delta \theta^\alpha =
\epsilon^\alpha\,.
\eea
It is called {\it left-chiral} ${\cal N}=1$ {\it superspace}.\\

\noindent In the basis $(x_L^m, \theta^\alpha, \bar\theta^{\dot\alpha})$ the chirality condition $(a)$ is reduced to
the {\it Grassmann Cauchy-Riemann} conditions:
\bea
\bar D_{\dot\alpha}\Phi_L
(x_L,\theta,\bar\theta)=0\; \Rightarrow \;\frac{\partial}{\partial
\bar\theta^{\dot\alpha}}\Phi_L = 0\; \Rightarrow \;\Phi_L =
\varphi_L(x_L,\theta)\,.
\eea

\subsection{Superfield actions}

Having superfields, one can construct out of them, as well as of their vector and covariant spinor derivatives,
scalar superfield Lagrangians. Any local product of superfields is again a superfield:
\bea
{\cal L} = {\cal L}(\Phi,
D_\alpha\Phi, \bar D_{\dot\alpha}\Phi, \partial_m\Phi, \ldots)\,,
\quad \delta {\cal L} = i\left(\epsilon^\alpha Q_\alpha +
\bar\epsilon_{\dot\alpha}\bar Q^{\dot\alpha}\right){\cal
L}\,.\nonumber
\eea

\noindent It is easy to see that the variation of the highest component in the $\theta$ expansion of any superfield is a total derivative. Then
one takes the highest component field in the $\theta$
expansion of the superfield Lagrangian and integrates it over
Minkowski space. It will be just an  action invariant under ${\cal N}=1$ supersymmetry!\\

\noindent A manifestly covariant way to write supersymmmetric actions is to use the Berezin integral. It is equivalent to differentiation
in Grassmann coordinates. In the considered case of ${\cal N}=1$ superspace it is defined by the rules
\bea
\int d^2 \theta \,(\theta)^2 = 1\,, \;\; \int d^2
\bar\theta\, (\bar\theta)^2 = 1\,, \;\; \int d^2\theta
d^2\bar\theta\, (\theta)^4 = 1\,,\;\; (\theta)^4 \equiv
(\theta)^2(\bar\theta)^2\,. \nonumber
\eea
Hence the Berezin
integral yields an efficient and manifestly supersymmetric way of
singling out the coefficients of the highest-order $\theta$ monomials in the superfield Lagrangians.\\

\noindent The simplest invariant action of chiral superfields producing the kinetic terms of the scalar multiplet is as follows
\bea
S_{kin}= \int d^4x d^4\theta\, \varphi(x_L, \theta)\bar\varphi (x_R, \bar\theta)\,, \quad x^m_R = \overline{(x_L^m)}
= x^m - i\theta \sigma^m\bar\theta\,.
\nonumber
\eea
After performing integration over Grassmann coordinates, one obtains
\bea
S \sim \int d^4 x \left(\partial^m\bar\phi \partial_m\phi -\frac{i}{2}\psi\sigma^m\partial_m\bar\psi + F\bar F \right).
\nonumber
\eea
\vspace{0.1cm}

\noindent The total Wess-Zumino model action is reproduced by adding, to this kinetic term, also potential superfield term
\bea
S_{pot} = \int d^4 x_L d^2\theta\, \left(\frac{g}{3}\varphi^3+ \frac{m}{2} \varphi^2\right) + \mbox{c.c.}\,.
\nonumber
\eea

\noindent This action is the only renormalizable action of the scalar ${\cal N}=1$ multiplet.
In principle, one can construct more general actions,
e.g., the action of K\"ahler sigma model and the generalized potential terms,
\bea
\tilde{S}_{kin} = \int d^4x d^4\theta\, K\left[\varphi(x_L, \theta), \bar\varphi (x_R,
\bar\theta)\right]\,, \quad \tilde{S}_{pot} = \int d^4 x_L d^2\theta\, P(\varphi) + \mbox{c.c.}\,.\nonumber
\eea
\vspace{0.2cm}

\noindent The multiplet with the superspin $Y =1/2$ is described by the gauge superfield $V(x,\theta,\bar\theta)$
possessing the gauge freedom
\bea
\delta V(x,\theta,\bar\theta) =i[\bar\lambda (x^m
-i\theta\sigma^m\bar\theta,\bar\theta) -\lambda (x^m
+i\theta\sigma^m\bar\theta,\theta)], \nonumber
\eea
where $\lambda(x_L, \theta)$ is an arbitrary chiral superfield parameter.\\

\noindent Using this freedom,  one can fix the so called Wess-Zumino gauge
\bea
V_{WZ}(x,\theta,\bar\theta) = 2\,\theta\sigma^m\bar\theta\, A_m(x)
+ 2i\bar\theta^2\theta^\alpha\, \psi_\alpha(x) -
2i\theta^2\bar\theta_{\dot\alpha}\,\bar\psi^{\dot\alpha}(x) +
\theta^2\bar\theta^2\, D(x)\;.\nonumber
\eea

\noindent Thus in the WZ gauge we are left with the irreducible set of fields forming the gauge (or vector) off-shell
supermultiplet: The gauge field $A_m(x)$, $A_m'(x) =  A_m + \partial_m\lambda(x)$, the fermionic field of gaugino
$\psi_\alpha(x)$, $\bar\psi_{\dot\alpha}(x)$ and the auxiliary field $D(x)$.\\

\noindent The invariant action is written as an integral over the chiral superspace
\bea
S^{N=1}_{{gauge}} = {1\over 16}\int d\zeta_L\; (W^\alpha
W_\alpha) + \mbox{c.c.}\,, \quad W_\alpha =
-\frac{1}{2} \bar{D}^2 D_{\alpha} V\,, \; \bar D_{\dot\alpha}W_\alpha =0\,.\nonumber
\eea

\noindent Everything is easily generalized to the non-abelian case. The corresponding component off-shell action reads
\bea
S=\int d^4x \, \mbox{Tr}\left[-{1\over 4}F^{mn}F_{mn}
-i\psi\sigma^m{\cal D}_m\bar\psi + {1\over 2}D^2\right].
\nonumber
\eea

\noindent What about superfield approach to higher ${\cal N}$ supersymmetries?  The difficulties arise because the relevant
superspaces contain too many $\theta$ coordinates and it is a very complicated problem to define the superfields which would
correctly describe the relevant irreps.\\

\noindent For ${\cal N}=2$, the off-shell gauge multiplet contains the vector gauge field $A_m(x)$, the complex
scalar physical field $\varphi(x)$, the $SU(2)$ doublet of Weyl fermions
$\psi_\alpha^i(x), \bar\psi_{\dot\alpha\,i}(x)$ and the auxiliary real $SU(2)$ triplet $D^{(ik)}(x)$.\\

\noindent There is no simple way to define ${\cal N}=2$ analog of the ${\cal N}=1$ gauge prepotential $V$  (unless we apply to
${\cal N}=2$ {\it harmonic superspace}).
However, one can define the appropriate covariant superfield strength $W$. In the abelian case,
it is defined by the off-shell constraints
\bea
(a)\,\bar D^i_{\dot\alpha} W =0\,, \quad (b)\,D^{\alpha i}D_{\alpha}^k W = \bar D_{\dot\alpha}^i\bar D^{\dot\alpha k}\bar  W\,,\nonumber
\eea
which, in particular, imply the Bianchi identity for the gauge field strength. The invariant action is an integral over chiral ${\cal N}=2$ superspace
\bea
S \sim \int d^4x_L d^4\theta\, W^2 + \mbox{c.c.}\,.\nonumber
\eea

\noindent What about maximally extended ${\cal N}=4$ super Yang-Mills? It has no superfield formulation with all
${\cal N}=4$ supersymmetries being manifest and off-shell. There is ${\cal N}=1$ superfield formulation
with one gauge superfield and three
chiral superfields; ${\cal N}=2$ formulation in terms of ${\cal N}=2$ gauge superfield and
one massless matter {\it hypermultiplet}. The latter possesses an off-shell formulation only in the
${\cal N}=2$ {\it harmonic superspace}. At last, exists a formulation with three manifest off-shell supersymmetries -
in ${\cal N}=3$ {\it harmonic superspace}. It involves gauge superfields only.

\section{Lecture V: Supersymmetric quantum mechanics}
\subsection{Supersymmetry in one dimension}
\noindent Quantum mechanics can be treated as one-dimensional field theory.  Correspondingly, the relevant supersymmetry can be understood as
the $d=1$ reduction of higher-dimensional Poincar\'e supersymmetry. More generally,  the ${\cal N}$-extended
$d=1$ ``Poincar\'e'' supersymmetry can be defined by the (anti)commutation relations
$$
\{Q^m, Q^n\} = 2 \delta^{mn}\,H\,, \quad [H, Q^m] = 0\,, \overline{Q^m} = Q^m\,,\; m = 1, \ldots {\cal N}\,.
$$
The associated systems are models of supersymmetric quantum mechanics (SQM)
with $H$ as the relevant Hamiltonian. The SQM models have a lot of applications in various physical and mathematical domains.\\

\noindent We will deal with the  simplest non-trivial ${\cal N} =2, d=1$ supersymmetry
$$
Q = \frac{1}{\sqrt{2}}(Q^1 + i Q^2)\,, \quad \bar Q = \frac{1}{\sqrt{2}}(Q^1 - i Q^2)\,,
$$
$$
\{Q, \bar Q\} = 2 H\,, \quad Q^2 = \bar Q^2 = 0, \quad [H, Q] = [H, \bar Q] = 0\,.
$$
It is also instructive to add the commutators with the generator $J$ of the group $O(2) \sim U(1)$
which is the automorphism group of the ${\cal N}=2$ superalgebra:
$$
[J, Q] = Q\,, \quad [J, \bar Q] = - \bar Q\,, \quad [H, J] = 0\,.
$$\\

\noindent ${\cal N} =2, d=1$ superspace is defined as:
\bea
{\cal M}^{(1|2)} = (t,\theta, \bar\theta)\,, \quad \delta \theta = \epsilon\,, \quad \delta \bar\theta = \bar\epsilon\,, \;\;
\delta t = i(\epsilon\bar\theta + \bar\epsilon\theta)\,. \nonumber
\eea

\noindent One can also define the ${\cal N}{=}\,2$ covariant spinor derivatives:
\begin{equation}
D = \partial_{\theta} -i\bar\theta\partial_{t}\,, \qquad \bar D = -\partial_{\bar\theta}
+i\theta\partial_{t}\,, \qquad \{D, \bar D \} = 2i \partial_{t}\,.\nonumber
\end{equation}

\noindent The simplest superfield is the real one, $\Phi(t, \theta, \bar\theta)$,
\bea
\Phi'(t', \theta', \bar\theta') = \Phi(t, \theta, \bar\theta)\; \Rightarrow \; \delta\Phi = -\delta t \partial_t\Phi
- \epsilon \partial_\theta\Phi - \bar\epsilon \partial_{\bar\theta}\Phi\,.\nonumber
\eea

\noindent On the component fields appearing in the $\theta$ expansion of $\Phi$,
\bea
\Phi(t, \theta, \bar\theta) =  x(t) + \theta \psi(t) -\bar\theta \bar\psi(t) +\theta \bar\theta y(t)\,,\nonumber
\eea
${\cal N}{=}\,2$ supersymmetry is realized as
\bea
\delta x = \bar\epsilon \bar\psi - \epsilon \psi\,, \; \delta\psi = \bar\epsilon(i\dot x - y)\,, \;
\delta\bar\psi = -\epsilon(i\dot x + y)\,, \; \delta y = i(\epsilon \dot\psi + \bar\epsilon \dot{\bar\psi})\,.\nonumber
\eea

\noindent The superfield $\Phi(t, \theta, \bar\theta)$ comprises the irreducible ${\cal N} =2, d=1$ multiplet
$({\bf 1, 2, 1})\,$. Other ${\cal N} =2, d=1$ multiplets exist as well, e.g., $({\bf 2, 2, 0})\,$, which is described by a chiral
${\cal N} =2, d=1$ superfield.\\

\noindent The simplest invariant superfield action containing interaction reads
\bea
S^{({\cal N}=2)} = \int dt d^2\theta \,\Big[\, \bar{D}
\Phi\,  {D} \Phi +  W(\Phi) \,\Big].\nonumber
\eea
Here $W(\Phi) $ is the superpotential. After integrating over Grassmann coordinates, we obtain
\bea
S^{({\cal N}=2)} = \int dt  \,\left[\, \dot x^2
- i\left(\dot{\bar\psi}\psi -{\bar\psi}\dot\psi\right) +y^2 + {y}\partial_x W(x)  + (\psi\bar\psi)\partial^2_x W(x)\,\right].\nonumber
\eea

\noindent The next step is to eliminate the auxiliary field $y$ by its algebraic equation of motion
$$
y = -\frac12 \partial_x W\,.
$$
The on-shell action is then
$$
S^{({\cal N}=2)} = \int dt  \,\left[\, \dot x^2
- i\left(\dot{\bar\psi}\psi  - {\bar\psi}\dot\psi\right) - \frac14 (\partial_xW)^2 +  (\psi\bar\psi)\partial^2_x W(x)\,\right].
$$\\

\noindent The action is invariant under the transformations
$$
\delta x = \bar\epsilon \bar\psi - \epsilon \psi\,, \;\delta\psi = \bar\epsilon(i\dot x + \frac12\partial_x W )\,, \;
\delta\bar\psi = -\epsilon(i\dot x - \frac12\partial_x W  )\,.
$$

\subsection{Hamiltonian formalism and quantization}

The quantum Hamiltonian obtained in a standard way from the canonical one reads
\bea
H = \frac{1}{4}\left[  \hat p{}^2+\,\left(\frac{d W}{d \hat x}\right)^2\,\right]
-\frac12\,\frac{d^2 W}{d \hat x^2}\left(\hat \psi\hat {\bar\psi} -\hat {\bar\psi}\hat \psi\right),\nonumber
\eea
where we have Weyl-ordered the fermionic term. The supercharges calculated by
the Noether procedure and then brought into the quantum form through passing to the operators are
\bea
Q = \hat \psi\left(\hat p+i\,\frac{d W}{d \hat x}\,\right),\qquad \bar Q = \hat {\bar\psi}\left(\hat p-i\,\frac{d W}{d \hat x}\,\right).\nonumber
\eea

\noindent The algebra of the basic quantum operators is
\bea
[ \hat x, \hat p]=i\,,\qquad \{ \hat\psi, \hat{\bar\psi}\}= {\textstyle\frac{1}{2}}\,. \nonumber
\eea
Using it, we can calculate the anticommutators of the quantum supercharges and check that they form ${\cal N}=2, d=1$ superalgebra
\bea
\{ Q, \bar Q\}  =2H\,,\qquad \{ Q,  Q\} =\{ \bar Q,  \bar Q\} = 0\,.
\eea

\noindent By the graded Jacobi identities, one also derives
$$
[Q, H] = [\bar Q, H] = 0\,.
$$

\noindent We use the standard realization for $\hat{p}$,
$
\hat{p} = \frac{1}{i}\frac{\partial}{\partial x}\,,
$
and the Pauli-matrix realization for the fermionic operators
$$
\hat \psi=\frac{1}{2\sqrt{2}}\left(\sigma_1 +i\sigma_2 \right), \quad \hat {\bar\psi}=\frac{1}{2\sqrt{2}}\left(\sigma_1 -i\sigma_2 \right),
\quad \hat \psi\hat {\bar\psi}-\hat {\bar\psi}\hat \psi = {\textstyle\frac{1}{2}}\,\sigma_3\,.
$$
Then the Hamiltonian and supercharges are represented by  $2\times 2$ matrices
\bea
&& H = \frac{1}{4}\left[ -{\partial_x}^2+\,\left(W_x\right)^2\,\right]\left(\begin{array}{cc}
                  1 & 0\\
                  0 & 1
                \end{array}\right)
-\frac14\,W_{xx}\,\left(\begin{array}{cc}
                  1 & 0\\
                  0 & -1
                \end{array}\right)\nn
&& Q = -\frac{i}{\sqrt{2}}\left(\begin{array}{cc}
                  0 & 1\\
                  0 & 0
                \end{array}\right)  \left(\partial_x - W_x\right),\quad
\bar Q = -\frac{i}{\sqrt{2}}\left(\begin{array}{cc}
                  0 & 0\\
                  1 & 0
                \end{array}\right) \left(\partial_x + W_x\right). \nonumber
\eea

\noindent Thus the wave functions form a doublet and, taking into account the conditions $[Q, H] = [\bar Q, H] = 0$,
the relevant matrix spectral problem is
\bea
H \left(\begin{array}{cc}
                  \psi_+\\
                  \psi_-
                \end{array}\right) = \lambda \left(\begin{array}{cc}
                  \psi_+\\
                  \psi_-
                \end{array}\right).\nonumber
\eea

\noindent It is equivalent to the two ordinary problems
\bea
H_\pm\psi_\pm  = \lambda\,\psi_\pm\,, \qquad H_\pm = -\frac14 (\partial_x \mp W_x)(\partial_x \pm W_x).\nonumber
\eea

\noindent Using the intertwining property
\bea
H_-(\partial_x + W_x) = (\partial_x + W_x)H_+\,, \quad H_+(\partial_x - W_x) = (\partial_x - W_x)H_-\,,\nonumber
\eea
it easy to show that the states
\bea
Q\left(\begin{array}{cc}
                 \psi_+\\
                  \psi_{-}
               \end{array}\right) = \left(\begin{array}{cc}
                -i(\partial_x -W_x) \psi_{-}\\
                 0
                \end{array}\right),
               \quad \bar{Q}\left(\begin{array}{cc}
                  \psi_+\\
                  \psi_{-}
               \end{array}\right)
               = \left(\begin{array}{cc}
                0\\
                  -i(\partial_x + W_x)\psi_{+}
               \end{array}\right)\nonumber
\eea
are the eigenfunctions of $H_+$ and $H_{-}$ with the same eigenvalue $\lambda$ as $\psi_+$ and $\psi_-$. Thus
we observe the double degeneracy of the spectrum. This double degeneracy is the most characteristic feature of the ${\cal N}=2$
supersymmetry in $d=1$ (and of any higher ${\cal N}$ supersymmetry in $d=1$).\\

\noindent In general, the Hilbert space of quantum states of ${\cal N}=2$ SQM is divided into the following three sectors
\bea
&&(a)\; \mbox{Ground state}:\; Q\Psi_0 = \bar{Q}\Psi_0 = H\Psi_0 = 0\,,\nn
&& (b)\; H\Psi_1 = E\Psi_1\,,\; Q\Psi_1 \neq 0\,, \;\bar{Q}\Psi_1 = 0\,, \nn
&& (c) \; H\Psi_2 = E\Psi_2\,, \;\bar{Q}\Psi_2 \neq 0\,, \; Q\Psi_2 = 0\,. \nonumber
\eea

\noindent Based on this consideration, one can conclude that many QM models with the double degeneracy
of the energy spectrum can be identified with some ${\cal N}=2$ SQM models.

\section{Summary}

  \begin{itemize}

  \item
  Supersymmetry between fermions and bosons is a new unusual concept in the mathematical physics. It allowed to construct a lot of new theories
  with remarkable and surprising features: supergravities, superstrings, superbranes, ${\cal N}=4$ super Yang-Mills theory (the first example of
  the ultraviolet-finite quantum field theory), etc. It also allowed to establish unexpected relations between these theories, e.g., the AdS/CFT
  (or ``gravity/gauge'') correspondence, AGT correspondence, etc.
  \vspace{0.2cm}

 \item It predicts new particles (superpartners) which still await their experimental discovery.
\vspace{0.2cm}

\item The natural approach to supersymmetric theories is the superfield methods.

\end{itemize}

\noindent For those who wish to get deeper insights into the subjects sketched in these lectures, I may recommend the text-books and
the review papers in the list of references below.

\section*{Acknowledgements} I thank the organizers of the International School in Tsakhkadzor and, personally, George Pogosyan
for inviting me to give these lectures and for the kind hospitality in Armenia.

\end{document}